\documentstyle[preprint,aps]{revtex}
\begin{document}
\draft
\title{Evidence for novel polaronic Fermi-liquid in doped oxides}

\author{A.S. Alexandrov$^{1,*}$, Guo-meng Zhao$^{2,**}$, H.
Keller$^{2}$, ~\\
B. Lorenz$^{3}$, Y. S. Wang$^{3}$, and C. W. Chu$^{3,4}$. }
\address
{$^{1}$Department of Physics, Loughborough University,
Loughborough LE11 3TU, UK\\
$^{2}$Physik-Institut der Universi\"at Z\"urich, CH-8057 Z\"urich,
Switzerland\\
$^{3}$Department of Physics and Texas Center for Superconductivity,
University of Houston, Houston, Texas 77204, USA\\
$^{4}$Lawrence Berkeley National Laboratory, 1 Cyclotron Road, Berkeley, Ca
94720, USA.}

\maketitle
%\vskip0.5truecm
%\noindent
\begin{abstract}

 The modern
multi-polaron theory  predicts a
{\it polaronic}  Fermi-liquid state, but such a state has not
been unambigously
confirmed by experiment so far.  Here we report theoretical and
experimental studies of the isotope effects on
the {\it low~temperature~kinetic} properties of the
doped ferromagnetic manganites. Our results provide clear evidence for a
novel polaronic  Fermi-liquid state in doped oxides.

\end{abstract}

\pacs{PACS numbers:74.20.Mn,74.20.-z,74.25.Jb}

\narrowtext
Over the last  decades, the very interesting phenomena such as
high-temperature superconductivity and colossal
magnetoresistance have been discovered in several doped
oxides (e.g., cuprates \cite{mul}, bismuthates \cite{Cava}, and
manganites \cite{Von,Jin}).
These doped oxides are characterized by significant carrier densities
($\geq 10^{21}$ cm$^{-3}$) and  low mobility  of the order or
even less than the
Mott-Ioffe-Regel
limit ($ea^{2}/\hbar \sim$ 1~cm $^{2}$/Vs, where $a$ is the lattice constant) at room temperature. The very
nature of
the low-temperature `metallic' state of these materials can be hardly
understood within the framework of the canonical theory of metals.
However,
 it has
been known for many decades starting from the
pioneering work by Landau \cite{lan} that the local lattice
deformation can transform  electrons into  `self-trapped' polarons in doped
ionic
insulators. Further extensive theoretical studies (for review see \cite{alemot,dev} have
shown that polarons behave like heavy particles, and can be mobile
with metallic conduction at sufficiently low temperatures. Under certain
conditions \cite{ale2} they form a polaronic Fermi liquid
with some properties being very different from ordinary metals. One of
them is the effective  mass of polarons.
It is well known that electrons can change their mass in solids due to the interactions with
ions, spins, and themselves. The renormalized
(effective) mass of electrons is
independent of the ion mass $M$ in ordinary metals where the Migdal
adiabatic approximation is believed to be valid.  However, if the
interactions between electrons and nuclear ions are strong and/or the
adiabatic approximation is not applied 
electrons  form polarons (quasiparticles dressed by lattice
distortions), so  their effective mass $m^{*}$ will depend on
$M$. This is because the polaron mass $m^{*}= m \exp (A/\omega)$
\cite{ale}, where $m$ is the band mass in the absence of the electron-phonon
interaction, $A$ is a constant, and $\omega$ is a characteristic
phonon frequency which depends on the masses of ions.  Hence, there
is a large isotope effect on the carrier mass in polaronic metals, in
contrast to the zero isotope effect in ordinary metals. 

The total exponent
of the isotope effect on $m^{*}$ is defined as $\alpha_{m^{*}}= \sum - d\ln
m^{*}/d\ln M_{i}$ ($M_{i}$ is the mass of the $i$th atom in a unit cell).
From this definition and the expression for the polaron mass $m^{*}$
mentioned
above, one readily finds
\begin{equation}
\alpha_{m^{*}} = -{1\over{2}}\ln(m^*/m).
\end{equation}
Interestingly, the same isotope exponent is predicted for the
inter-site bound pairs of polarons (i.e., bipolarons) \cite{ale,tru}.
A large isotope effect on the effective mass of supercarriers has been
observed in the high-temperature cuprate superconductors
\cite{ZhaoPRB95,ZhaoNature97,Hofer}, which
suggests that polarons and/or bipolarons should be involved in
high-temperature superconductivity. Furthermore, in doped
manganites, the polaronic nature of
the carriers in
the high-temperature paramagnetic insulating state has been shown by
several independent experimental
results such as the sign
anomaly of the Hall effect \cite{emi}, the Arrhenius behavior of the drift
and Hall
mobilities \cite{emi}, the giant
oxygen-isotope shift of the Curie temperature $T_{C}$ \cite{ZhaoNature96},
and the electron paramagnetic resonance
linewidth which is proportional to the conductivity \cite{Sheng}.

On the other hand, the very nature of the charge carriers in the
low-temperature metallic state of doped manganites has not been
understood. In a theory of colossal magnetoresistance for
ferromagnetic manganites \cite{alebra} polarons are considered as the
carriers even in the low-temperature metallic state, while others
\cite{rev,good} believe that polaronic effects are not important at
low temperatures. In particular, it was shown that, in low $T_{C}$
polycrystalline manganites ($T_{C}<$120K), the resistivity and
$T_{C}$ are extremely sensitive to the external pressure \cite{good}
and chemical pressure \cite{Uehara} while the
thermopower is small and nearly pressure independent \cite{good}. This
unusual behavior in the low $T_{C}$ materials was taken
as evidence for a vibronic electronic state  \cite{good}, or may be
explained in terms of a percolative conduction
mechanism \cite{Uehara}. In contrast, from isotope-effect
studies \cite{Zhou}, it was concluded
that polaronic effects may be involved in the ferromagnetic
metallic state of the low $T_{C}$ compounds. Moreover, in high $T_{C}$
materials, it is even less clear
whether polaronic effects are important in the low temperature metallic
state.

Theoretically it has been demonstrated
that the coherent motion of small polarons can lead to metallic
conduction at low temperatures \cite{alemot,fir,ale2}. However, there have
been no
convincing and clear experiments to confirm this theoretical prediction.
In order to provide decisive evidence for the
`metallic' polaronic liquid in doped ferromagnetic manganites, it is
essential to
study the isotope effects on the low-temperature
kinetic and thermodynamic properties. These properties are dominated by the
impurity and
electron-electron scatterings, so the only source of the isotope effect
might be the polaron mass dependence on the ion mass. For example,
replacing $^{16}$O with $^{18}$O will increase
the polaron mass, leading to the changes in the residual
resistivity, and in other kinetic and thermodynamic properties.
Here we provide clear evidence for the
polaronic liquid in doped ferromagnetic manganites at low temperatures
from theoretical and experimental studies of the isotope effects
on kinetic and thermodynamic properties.

We first define and calculate three exponents:
$\beta_C=d\ln C_{el}/d\ln m^*$,
$\beta_{\rho}= d\ln\rho_{o}/d \ln m^*$, and $\beta_S = d\ln S/d\ln m^*$ in the
polaronic
liquid. Here $C_{el}$ is the electronic specific-heat at low temperatures,
$\rho_{o}$ is the residual resistivity, and $S$ is the
low-temperature thermopower. The above definition is equivalent to
$C_{el} \propto (m^{*})^{\beta_C}$, $\rho_{o} \propto
(m^{*})^{\beta_\rho}$, and
$S \propto (m^{*})^{\beta_S}$. We assume that the Fermi level (chemical
potential)  lies above the mobility
edge, so  polarons
propagate in the Bloch states as heavy fermions \cite{fir,ale2}
scattered by the short-range impurity potential.  Since doped holes in
manganites mainly reside on the oxygen orbitals \cite{Saitoh,Ju}, their
bands are almost one-dimensional (1D) due
to a large anisotropy in the hopping integrals (i.e., $t_{pp\pi} =
-(1/4)t_{pp\sigma}$ \cite{alemot}). The polaronic mass enhancent
factor could be larger along the directions with a much smaller hopping
integral \cite{kor}. This will increase the anisotropy further so that the
polaron bands become more 1D-like. Then the
polaronic
propagator, $G(k,E)=[E-k^2/2m^*-\Sigma(k,E)]^{-1}$  is found
analytically with the self-energy, which is momentum ($k$)-independent in
the non-crossing approximation,
\begin{eqnarray}
\sigma(\epsilon)&=& {\epsilon\over{3}}
-\left
({1-i3^{1/2}\over{2}}\right)\left[{1\over{16}}+{\epsilon^{3}\over{27}}+\left ({1
\over{256}}+
{\epsilon^{3}\over{216}}\right )^{1/2}\right ]^{1/3}\cr
&-&\left ({1+i3^{1/2}\over{2}}\right )\left
[{1\over{16}}+{\epsilon^{3}\over{27}}-\left ({1\over{256}}+
{\epsilon^{3}\over{216}}\right )^{1/2}\right ]^{1/3}.
\end{eqnarray}
Here we introduce
 a dimensionless energy,
  $\epsilon = E/\epsilon_{0}$, and the self-energy
 $\sigma(\epsilon)=\Sigma(E)/\epsilon_0$ using the scattering rate
 $\epsilon_{0}=(D^{2}m^{*}/\hbar^{2})^{1/3}$ as  the energy unit.
The constant $D$ is the second moment of the Gaussian white-noise potential,
$D=2(n_{im}v^{2}/a^{2})$, where  $n_{im}$ is the
impurity density, and $v$ is the coefficient of
the $\delta$-function
impurity potential. This self-energy is almost exact above
 the mobility edge as shown with the exact solution for a
 one-dimensional particle  in a random white-noise potential
 \cite{fri}.

The specific heat is proportional to  the density of states (DOS)
 at the (dimensionless) Fermi
 level, $\mu=E_{F}/\epsilon_{0}$. Assuming three degenerate spin-polarized
 oxygen bands, we find
\begin{equation}
{C_{el}\over{T}} ={2 \pi k_{B}^{2} \over{\hbar a^{2}}}\left
({m^*\over{\epsilon_0}}\right)^{1/2} g(\mu)
\end{equation}
with  $g(\epsilon)= \Im \sigma(\epsilon)$. The residual  conductivity
$\rho_{o}^{-1}$ and the low-temperature
thermopower $S$ are found with the Kubo
formalism \cite{kub} as
\begin{equation}
\rho_{o}^{-1}= {e^2\over{2\pi^2(m^*)^2}}\int^{\infty}_{-\infty}dk k^2
\left[\Im G(k,\mu)\right]^{2},
\end{equation}
and
\begin{equation}
S=-{\pi^2k_B^2T\over{3e\epsilon_{0}}}{d\ln \rho_{o}\over{d\mu}}.
\end{equation}
If there coexist electron-like and hole-like  polarons, Eq.~5 must be
multiplied by a factor $f_{S}$ with $|f_{S}|\leq 1$. We would like to mention
that Eq.~4 should be underestimated due to the
fact that the bands are not strictly one-dimensional and that
the transport relaxation rate is generally smaller than $\epsilon_{0}$.
Calculating the integral in Eq.~4 leads to
\begin{equation}
\rho_{o}^{-1}= {e^2\over{2a^{2}\pi (m^* \epsilon_0)^{1/2}}}F(\mu),
\end{equation}
where
\begin{equation}
F(\mu)={\tilde{\mu}[\tilde{\mu}+(\tilde{\mu}^2+g^2)^{1/2}]+g^2/2\over
{g[(\tilde{\mu}^2+g^2)(\tilde{\mu}+(\tilde{\mu}^2+g^2)^{1/2})]^{1/2}}},
\end{equation}
$\tilde{\mu}=\mu- \Re \sigma(\mu)$ and $g=g(\mu)$.

These expressions allow us to calculate the  exponents: $\beta_C$,
$\beta_{\rho}$, and $\beta_S$. The number of carriers does not
depend on the isotope substitution \cite{Zhaocond}, so that  the  chemical
potential itself
depends on the polaron mass with the exponent
\begin{equation}
{d \mu\over{d\ln m^*}}=-{2N\over{3g}},
\end{equation}
where $N=\int_{-\infty}^{\mu} g(\epsilon) d\epsilon$ is
proportional to the total number of states under the Fermi level.
Keeping that in mind we finally obtain
\begin{equation}
\beta_C= {1\over{3}}\left(1-{2Ng'\over{g^2}}\right),
\end{equation}
\begin{equation}
\beta_{\rho}= {2\over{3}}\left(1+{N F'\over{g F}}\right),
\end{equation}
and
\begin{equation}
\beta_S= {1\over{3}}\left({2N (F'/F-F''/F')\over{g}}-1\right).
\end{equation}
The prime and the double primes mean the first and the second
derivative with respect to $\mu$. The numerical results for the exponents
are shown in Fig.~1 as a function of the Fermi level
$E_{F}/\epsilon_0$. To compare with the experimentally measured
oxygen-isotope isotope exponents one should multiply $\beta_C, \beta_{\rho},
\beta_S$  by
the oxygen-isotope exponent $\alpha^{O}_{m^{*}}$.
Because of the logarithmic dependence on the
mass ratio (see Eq.~1),  the magnitude of the exponent $\alpha_{m^{*}}$ is
expected to
be of the order
of unity. At large filling, i.e.,  $\mu >>1$, the theory predicts
sizable isotope effects on all thermodynamic and kinetic properties
with the exponents:
\begin{equation}
\beta_C =1;  \beta_{\rho}=2;  \beta_S= 1.
\end{equation}
This is precisely what one expects  from the simple Born approximation for
the impurity
scattering.  The ratio remains the same for any dimension of the
polaron spectrum in this doping regime. Interestingly, if the doping
is not so high, $\mu \leq 1$, the exponent for the thermopower is very
small (see Fig. 1), while the exponent
for the residual resistivity remains larger than unity.
For $\mu = 0.25$ and  $\alpha^{O}_{m^{*}}\simeq \alpha_{m^{*}}\simeq
1.1$,
\begin{equation}
\beta_C \simeq 0.45; \beta_{\rho} \simeq 1.45; \beta_S \simeq -0.03,
\end{equation}
and
\begin{equation}
\alpha^{O}_C \simeq -0.5; \alpha^{O}_{\rho} \simeq -1.6; \alpha^{O}_S
\simeq 0.0,
\end{equation}
i.e., negligible isotope effect on $S$, and
more than  3 times lower isotope effect on $C_{el}$ than on $\rho_{o}$.
Applying
the exact density of states in the random potential \cite{fri},
we find even a somewhat lower exponent $\beta_C$ for $\mu
\leq 0$ than that in Fig.~1. All these findings are in sharp contrast
with those in ordinary metals where $\alpha_{m^{*}}= 0$, and there are no
isotope effects on $C_{el},\rho_{o}$ and $S$.

The choice of $\mu = 0.25$ and $\alpha^{O}_{m^{*}} = -1.1$ is consistent
with the experimental results for Nd$_{0.67}$Sr$_{0.33}$MnO$_{3}$.
Using the measured bare plasma frequency $\hbar\Omega_{p}$ $\simeq$ 3.3 eV
\cite{ZhaoPRL2000} and the total carrier density per cell of
0.33,  we obtain the bare hopping integral $
t = t_{pp\sigma} = \hbar^{2}/2a^{2}m$ = 0.9 eV, which is very close to the
value (0.87 eV) calculated for La$_{2}$CuO$_{4}$ \cite{Mac}. From Eq.~1 and
$\alpha_{m^{*}} \simeq \alpha^{O}_{m^{*}} \simeq -1.1$, we get the
effective hopping integral $t^{*}=\hbar^{2}/2a^{2}m^{*}$ $\simeq$ 100 meV.
By setting $\mu$ =
0.25, $t^{*}$ = 100 meV, $\epsilon_{0}$ = 27 meV, we calculates $C_{el}/T$
= 26
mJ/moleK$^{2}$ and $\rho_{o}$ = 185 $\mu\Omega$ cm by using Eq.~3 and
Eq.~6, respectively.  These values are
in remarkably good agreement with the measured values: $C_{el}/T$ =
25(3) mJ/moleK$^{2}$ \cite{Gordon}, and $\rho_{o}$ $\simeq$ 170 $\mu\Omega$
cm for the best single crystals \cite{Dai,Sawaki}. Furthermore, the value
of  $\epsilon_{0}$
is in good agreement with the measured transport relaxation rate
($\sim$24
meV)  for
La$_{0.7}$Ca$_{0.3}$MnO$_{3}$ \cite{Simpson}. This justifies the choice of
$\mu = 0.25$ and $\alpha^{O}_{m^{*}} = -1.1$ for
Nd$_{0.67}$Sr$_{0.33}$MnO$_{3}$. For La$_{0.75}$Ca$_{0.25}$MnO$_{3}$,
we calculate $C_{el}/T$ = 7.5 mJ/moleK$^{2}$ and $\rho_{o}$ =
150 $\mu\Omega$ cm by setting $\mu = 0.70$ and $\alpha^{O}_{m^{*}} = -
0.7$. The measured $\rho_{o}$ for La$_{0.75}$Ca$_{0.25}$MnO$_{3}$ is
120-150 $\mu\Omega$ cm (see Fig.~2a and discussion in
Ref.~\cite{Zhaocond}). From our preliminary specific-heat data
for La$_{0.75}$Ca$_{0.25}$MnO$_{3}$, we estimate $C_{el}/T$ $\simeq$ 7
mJ/moleK$^{2}$, which is slightly lower than the one ($\sim$8
mJ/moleK$^{2}$) for
La$_{0.8}$Ca$_{0.2}$MnO$_{3}$ \cite{Hamilton}. With $\mu = 0.70$ and
$\alpha^{O}_{m^{*}} = -
0.7$, we have
\begin{equation}
\alpha^{O}_C \simeq -0.5; \alpha^{O}_{\rho} \simeq -1.2; \alpha^{O}_S \simeq
-0.17,
\end{equation}

Experimentally, the oxygen-isotope effect on the intrinsic low-temperature
resistivity
has been studied \cite{Zhaocond} in high-quality thin films of
La$_{0.75}$Ca$_{0.25}$MnO$_{3}$ and Nd$_{0.7}$Sr$_{0.3}$MnO$_{3}$.
Our thin films were thick (2000 $\AA$), and were annealed at 940$^o$C for 10 hours, so no strains could  remain in
the films. We did back-exchange, and we knew  precisely the
oxygen-isotope enrichment (more details can be found in Ref. \cite{Zhaocond,ZhaoPRB99}).
In Fig.~2 we plot the low-temperature resistivity of the oxygen-isotope
exchanged films of (a)  La$_{0.75}$Ca$_{0.25}$MnO$_{3}$;
(b) Nd$_{0.7}$Sr$_{0.3}$MnO$_{3}$. In both cases, the
residual resistivity $\rho_{o}$ for the $^{18}$O samples is larger than
for the $^{16}$O samples by about 15$\%$. The observed large oxygen-isotope
effect on $\rho_{o}$ is consistent with the above theoretical prediction for
polaronic metals. From the definition of the oxygen-isotope
exponent: $\alpha_{\rho}^{O}$ = $-d\ln \rho_{o}/d\ln M_{O}$ ($M_{O}$
is the oxygen mass), we obtain
$\alpha_{\rho}^{O} = -1.4(2)$ for Nd$_{0.7}$Sr$_{0.3}$MnO$_{3}$,
and $\alpha_{\rho}^{O} = -1.2(2)$ for La$_{0.75}$Ca$_{0.25}$MnO$_{3}$.
These values are in excellent agreement with Eq.~14 and Eq.~15,
respectively.

The quantitative agreement between the resistivity data and the
theory indicates that the low-temperature metallic state in doped
manganites is a polaronic Fermi-liquid. To further confirm this,
one should experimentally demonstrate that $\alpha^{O}_S$ is
negligible and that $\alpha^{O}_C$ is rather small in these materials.
Our preliminary results of the oxygen-isotope effect on
the low-temperature specific heat in La$_{1-x}$Ca$_{x}$MnO$_{3}$
are consistent with Eq.~15 within the experimental uncertainty. The
results will be published elsewhere. Here we show the data of the thermopower
$S$ at low temperatures for the isotope-exchanged
La$_{1-x}$Ca$_{x}$MnO$_{3}$. The detailed measurement technique was described
in Ref.~\cite{Heilman}. Fig.~3a shows the thermopower $S$ below 60 K
for the $^{16}$O and  $^{18}$O samples of La$_{1-x}$Ca$_{x}$MnO$_{3}$
with $x$ = 0.20 and 0.25.  The isotope-exchanged samples are the same
as those reported in \cite{ZhaoPRB99}. It is apparent that
the oxygen-isotope effect on $S$ is negligible below 20 K for both
compositions ($\alpha_{S}^{O} \simeq 0.0$), as seen more clearly from Fig.~3b.
This is in sharp contrast to
the large oxygen-isotope shift of the
Curie temperature T$_{C}$, which is 21 K for $x$ = 0.20 and 14.5 K for $x$ =
0.25 \cite{ZhaoPRB99}. The fact that the slope of the linear part in
S (see Fig.~3) is very sensitive to the doping or the carrier
concentration and that the $^{16}$O and  $^{18}$O samples have the
same slope implies that the carrier concentrations for two isotope
samples of a fixed $x$ should be nearly the same, in agreement with
the previous results \cite{Zhaocond,ZhaoPRB99}.
The negligible $\alpha^{O}_S$ and the
substantial $\alpha^{O}_{\rho}$ $\sim$ -1.3 observed in these samples are in
quantitative agreement with the theoretical prediction (Eq.~14 and
Eq.~15). We are not aware of any other theoretical models which can
quantitatively explain the experimental values of several physical
quantities, such as electronic specific heat, residual resistivity,
scattering rate, and three different oxygen-isotope exponents. Therefore,
the present theoretical and
experimental studies of the low-temperature kinetic and thermodynamic
properties
provide strong evidence for the existence of novel polaronic Fermi-liquid
in doped
metallic oxides (e.g., magnetoresistive manganites and high-temperature
superconductive cuprates).
~\\
~\\
{\bf Acknowlegdment:} We would like to thank K. A. M\"uller for
valuable comments. The work at Z\"urich was supported by the Swiss
National Science Foundataion. The work at Houston was supported in
part by the US National Science Foundation, and work at Berkeley by the
US Department of Energy.

\newpage
{\bf Figure captions}

\vspace{0.5cm}

Fig.1. The numerical results for three exponents:
$\beta_\rho$, $\beta_C$, and $\beta_S$  as a
        function of the Fermi level $\mu \equiv$ $E_{F}/\epsilon_0$. See
        text for the definition of the exponents.

Fig.2. The low-temperature resistivity of the oxygen-isotope
exchanged
        films of (a) La$_{0.75}$Ca$_{0.25}$MnO$_{3}$; (b)
Nd$_{0.7}$Sr$_{0.3}$MnO$_{3}$. In both cases, the
residual resistivity $\rho_{o}$ for the $^{18}$O samples is larger than
for the $^{16}$O samples by  15$\%$\cite{Zhaocond}. The oxygen-isotope exponent $\alpha^{O}_{\rho}$
was calculated from the definition: $\alpha_{\rho}^{O}$ = $-d\ln
\rho_{o}/d\ln M_{O}$ ($M_{O}$
is the oxygen mass). 

Fig.3. The temperature dependence of the thermopower $S$
        of the $^{16}$O and
        $^{18}$O samples of La$_{1-x}$Ca$_{x}$MnO$_{3}$  below 60 K (Fig.~3a)
        and below 20 K (Fig.~3b). The solid and dash lines in Fig.~3b are the
     fitted curves by $S = S_{o}+ BT$ with a constraint that the
     absolute value of $S_{o}$ should be less than the systematic
     measurement error 0.1 $\mu$V/K (Ref.~\cite{Heilman}). The magnitudes
of the exponent
        $\alpha^{O}_{S}$ were estimated from the values of the
        fitting parameter $B$.


\begin{references}
\bibitem[*]{email} Email: a.s.alexandrov@lboro.ac.uk
\bibitem[**]{email} Email: zhao@physik.unizh.ch
\bibitem{mul}J. G. Bednorz $\&$ K. A.   M\"uller,  Z. Phys. B ${\bf
64}$, 189-193 (1986).
\bibitem{Cava} R. J. Cava, B.  Batlogg, J. J.  Krajewski, R.  Farrow, 
L. W. Rupp Jr., A. E.  White, K.  Short, W. F.  Peck,  $\&$
T. Kometani,  Nature (London)
\textbf{332},
814-816 (1988).
\bibitem{Von} R. M. von~Helmolt, J.  Wecker, B. Holzapfel, L.  Schultz,
 $\&$ K.  Samwer, 
Phys. Rev. Lett. \textbf{71}, 2331-2333 (1993).
\bibitem{Jin} S. Jin {\em et al.},  Science \textbf{264}, 413-415 (1994).
\bibitem{lan}
L. D. Landau, J. Phys. (USSR) ${\bf 3}$, 664 (1933).
\bibitem{alemot}
A. S. Alexandrov $\&$ N. F.  Mott,  {\em Polarons and Bipolarons}
(World Scientific, Singapore, 1995).
\bibitem{dev}
J. T. Devreese,   in
{\it Encyclopedia of Applied Physics}, vol. 14, p. 383, VCH Publishers
(1996).
\bibitem{ale2}
A. S. Alexandrov,  Phys. Rev. B {\bf 61}, 12315-12327 (2000).
\bibitem{ale}
A. S. Alexandrov,   Phys. Rev. B {\bf 46}, 14932-14935 (1992).
\bibitem{tru}
J. Bonca, T.  Katrasnik,  $\&$ S.  Trugman,  Phys. Rev. Lett. {\bf 84},
3153-3156 (2000).
\bibitem{ZhaoPRB95} G. M. Zhao,  $\&$ D. E.  Morris, 
 Phys. Rev. B {\bf 51}, R16 487-R16 490
(1995).
\bibitem{ZhaoNature97}
G. M. Zhao, M. B.  Hunt, H.  Keller, $\&$
K. A. M\"uller,  Nature (London)
\textbf{385},
236-238 (1997).
\bibitem{Hofer}J. Hofer, K.  Conder, T.  Sasagawa, G. M. Zhao, M.  Willemin,
H. Keller,  $\&$ K.  Kishio,   Phys. Rev. Lett. {\bf 84},
4192-4195 (2000).
\bibitem{emi}
M. Jaime,  {\em et al.}, Phys. Rev. Lett. {\bf 78}, 951-954 (1997).
\bibitem{ZhaoNature96} G. M. Zhao, K.  Conder, H. Keller,  $\&$
K. A. M\"uller,   Nature (London) \textbf{381},
676-678 (1996).
\bibitem{Sheng} A. Shengelaya, G. M.  Zhao, H. Keller, 
 K. A. M\"uller, $\&$ B. I. Kochelaev,  Phys. Rev. B \textbf{61},
5888-5890 (2000).
\bibitem{alebra} A. S. Alexandrov $\&$ A. M.  Bratkovsky,   Phys. Rev.
Lett. \textbf{82},
141-144 (1999).
\bibitem{rev}
for a recent review, see Y. Tokura $\&$ N.  Nagaosa,  Science {\bf 288},
462-468 (2000).
\bibitem{good} J. S. Zhou,  W. Archibald,  $\&$ J. B.  Goodenough,   Nature
(London) \textbf{381},
770-772 (1996).
\bibitem{Uehara} M. Uehara, C.  Mori, C. H. Chen, $\&$ S. W. Cheong,   Nature
(London) \textbf{399},
560-563 (1999).
\bibitem{Zhou} J. S. Zhou, $\&$ J. B.  Goodenough,   Phys. Rev. Lett.
\textbf{80}, 2665-2668 (1998).
\bibitem{fir} I. G. Lang $\&$ Yu. A.  Firsov,   Sov. Phys. -JETP
\textbf{16}, 1301-1312 (1963).
\bibitem{Saitoh} T. Saitoh, A. E.  Bocquet, T.  Mizokawa, 
H. Namatame, A.  Fujimori, M.  Abbate, Y.  Takeda,  $\&$ M.  Takano,  Phys.
Rev. B \textbf{51}, 13942-13951 (1995).
\bibitem{Ju} H. L. Ju, H. C.  Sohn, $\&$ K. M. Krishnan,   Phys.
Rev. Lett.
\textbf{79}, 3230-3233 (1997).
\bibitem{kor}
P. E. Kornilovitch,  Phys. Rev. B {\bf 59}, 13531-13535 (1999).
\bibitem{fri}
H. L. Frisch $\&$ S. P.  Lloyd,  Phys. Rev. ${\bf 120}$, 1175-1189 (1960).
\bibitem{kub}
R. Kubo, H. Hasegava,   $\&$ N. J.  Hashitsume,  Phys. Soc. Jap.{\bf 14},
56 (1959).
\bibitem{Zhaocond} G. M. Zhao, D. J.  Kang, W.  Prellier, M.  Rajeswari,
H. Keller, T.  Venkatesan, R. L. Greene,  cond-mat/0008029.
\bibitem{ZhaoPRL2000} G. M. Zhao, V.  Smolyaninova, W.   Prellier,   $\&$
H. Keller,  Phys. Rev. Lett. \textbf{84}, 6086-6089 (2000).
\bibitem{Mac} A. K. McMahan, R. M. Martin,  $\&$
S. Satpathy,  Phys. Rev. B
\textbf{38}, 6650-6666 (1988).
\bibitem{Gordon}J. E. Gordon, R. A. Fisher, Y. X.  Jia, N. E. Phillips,
S. F.  Reklis, D. A. Wright,  $\&$ A.  Zettl,  Phys.
Rev. B \textbf{59}, 127-130 (1999).
\bibitem{Dai} P. Dai, H. Y. Hwang, J.  Zhang, J. A.  Fernandez-Baca, 
S. W. Cheong, C.  Kloc, Y.  Tomioka,  $\&$ Y. Tokura,   Phys. Rev.
B \textbf{61}, 9553-9557 (2000).
\bibitem{Sawaki} Y. Sawaki, K.  Takenaka, A. Osuka, R. Shiozaki,  $\&$
S. Sugai, Phys. Rev.
B \textbf{61}, 11 588-11 593 (2000).
\bibitem{Simpson} J. R. Simpson, H. D.  Drew, V. N.  Smolyninova, 
R. L. Greene, M. C. Robson, A.  Biswas,  $\&$ M.  Rajeswari,  Phys. Rev. B
\textbf{60},
R16 263-R16 266 (1999).
\bibitem{Hamilton}J. J. Halmilton, E. L.  Keatley, H. L. Ju, 
A. K. Raychaudhuri, V. N.  Smolyaninova,  $\&$
R. L. Greene,  Phys. Rev. B \textbf{54}, 14926-14929 (1996).
\bibitem{Heilman} A. K. Heilman {\em et al.},  Phys. Rev. B
\textbf{61}, 8950-8954 (1999).
\bibitem{ZhaoPRB99} G. M. Zhao, K. Conder, H. Keller,  $\&$
K. A. M\"uller,   Phys. Rev. B
\textbf{60}, 11914-11917 (1999).
\end{references}
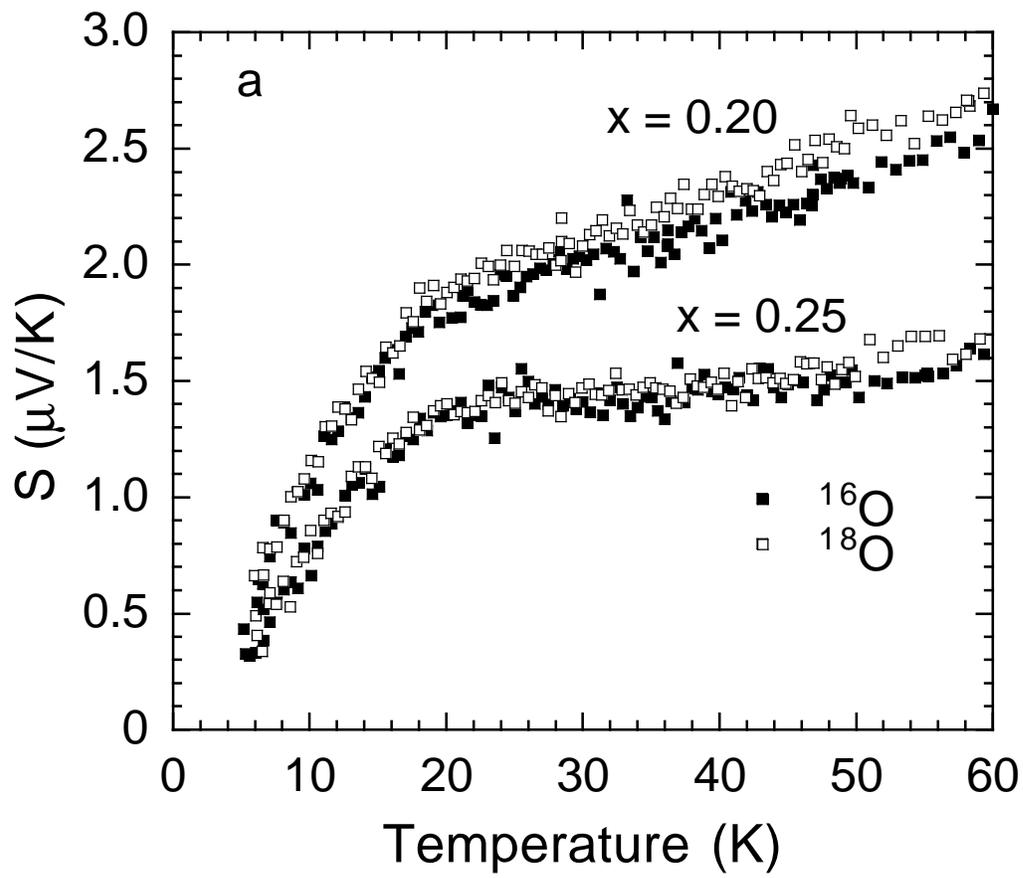
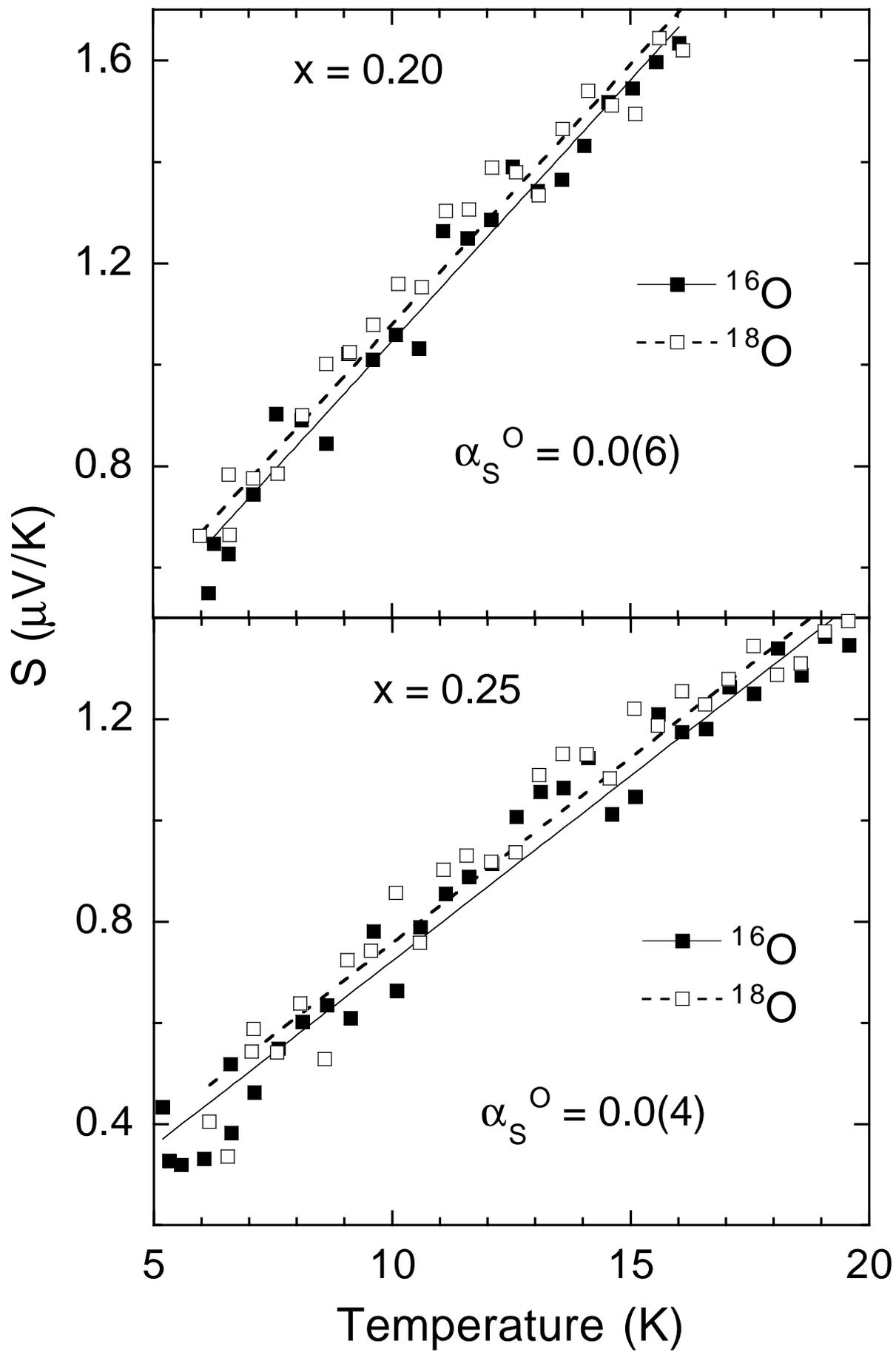
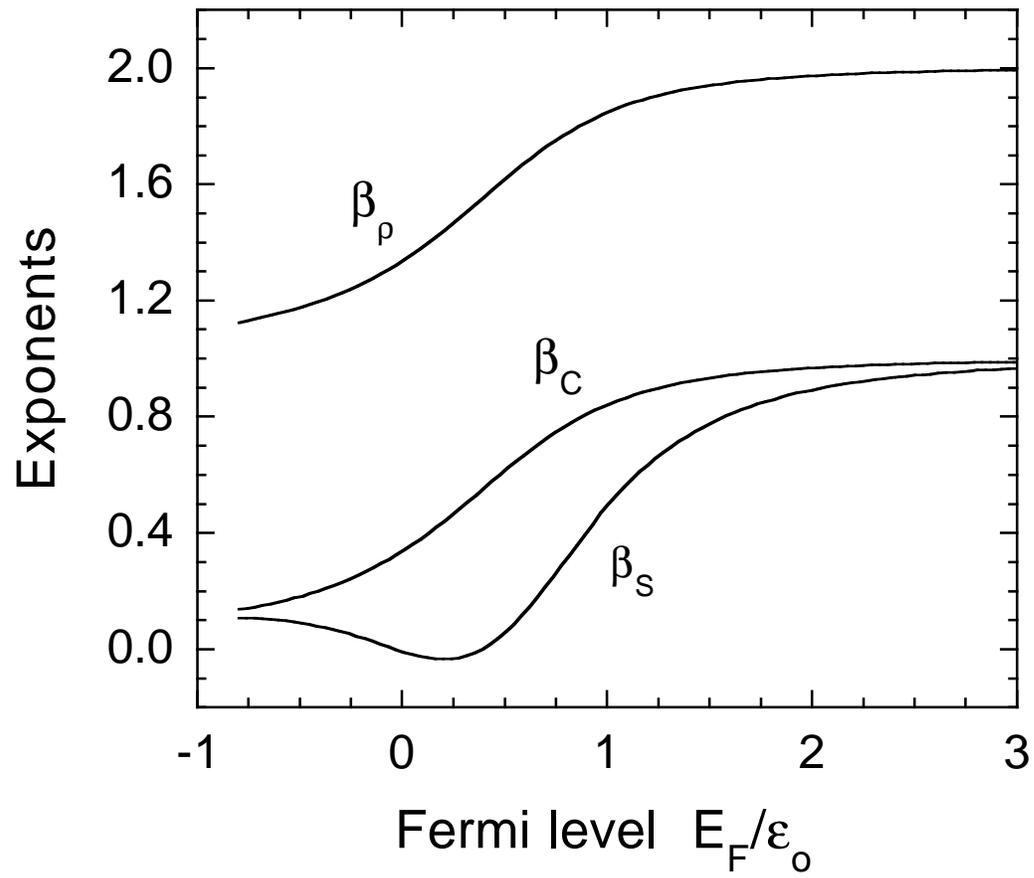
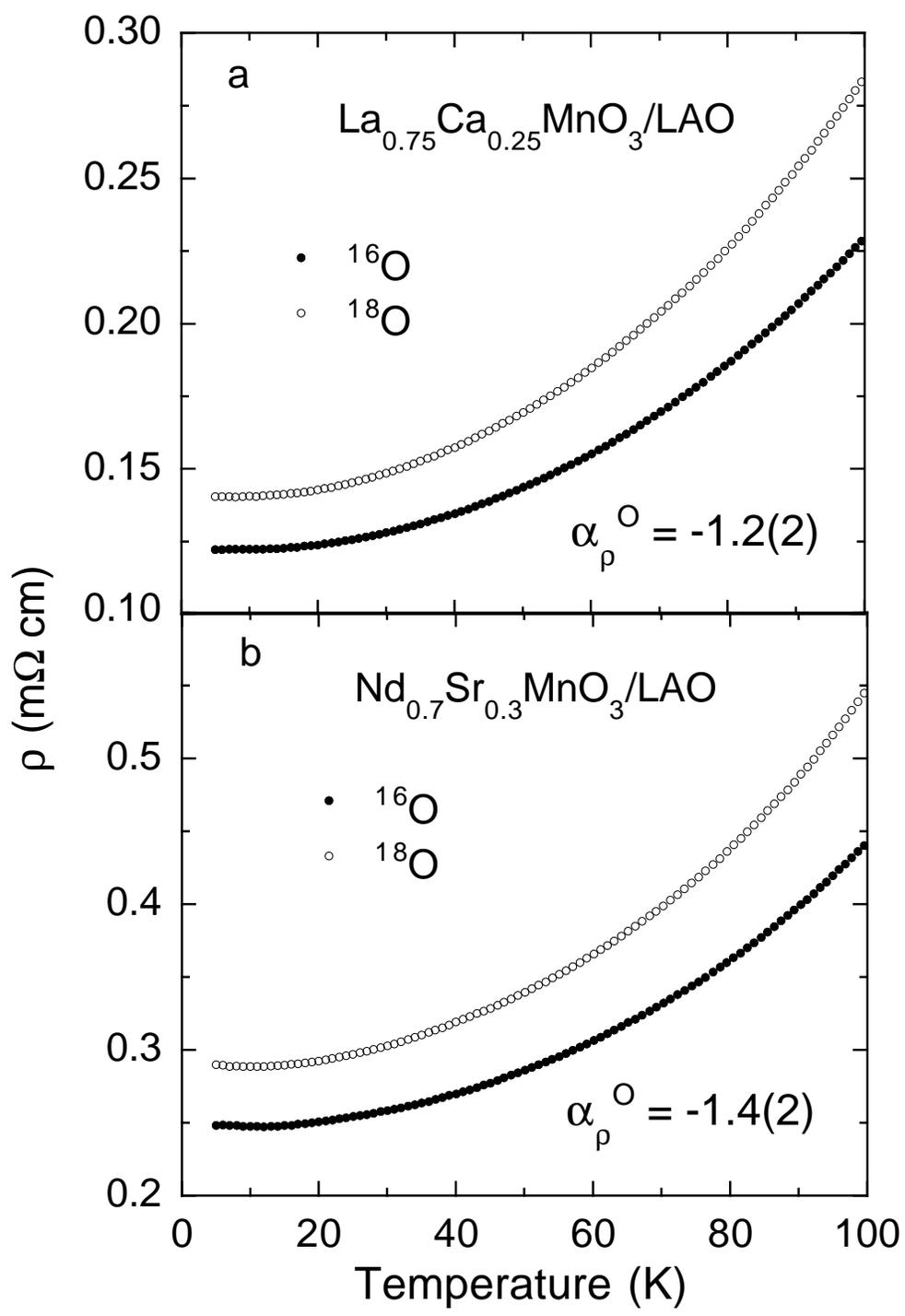
\end{document}